\documentstyle[prb,eqsecnum,aps]{revtex}
\begin{document}

\draft

\title{Been Down So Long it Looks Like Up to Me: A Unified Derivation of Conjugate
Gradient and Variable Metric Minimization}
\author{R. A. Hyman, Bridget Doporcyk, John Tetzlaff}
\address{Department of Physics, DePaul University, Chicago, IL
60614-3504}
\date{\today}
\maketitle

\begin{abstract}
Simple derivations, at a level appropriate for an undergraduate
computational physics course, of the most popular methods for
finding the minimum of a function of many variables are presented
in a unified manner in the context of a general optimization
scheme that emphasizes their essential similarities.  The
derivations in this paper encompass the conjugate gradient methods
with and without conditioning, and the variable metric methods.
The common variants of these methods including Fletcher-Reeves,
Polak-Ribiere, Davidon-Fletcher-Powell, and
Broyden-Fletcher-Goldfarb-Shanno are described and motivated.
\end{abstract}

\section{introduction}
You are near the top of a long valley and are meeting your friend
at the valley's lowest point for a picnic. You could probably see
the lowest point just by looking around and could walk right to
it. To make things challenging you have decided to find the
minimum in the same way that a computer would. First you
blind-fold yourself, since the information that you get from your
eyes, which includes the heights of a huge area around your
location, would be far too costly for a computer to calculate. To
find the lowest point in the valley while blindfolded you would
probably walk down hill, continuously changing your orientation to
be always pointing in the direction of steepest descent, which you
can determine with your feet alone. The path you would take is the
same path that a ball would take if velocity was proportional to
the force as in Aristotilian dynamics.  Regardless, this method is
also very costly for a computer, since it requires determination
of the steepest descent direction at every point along one's
trajectory. So to make your path more closely match what a
computer would do you have procured a magic scooter. You control
the initial orientation of the scooter, hop on, and away you go.
The scooter takes a straight line path and stops at the point
where to go any farther would take you up hill\cite{scooter}. This
is almost never the lowest point in the valley.  When the scooter
stops you can change its orientation and take another ride. This
curious way of going downhill actually resembles the procedure
many computer algorithms use to find the minimum of a function of
many variables. Different methods vary only in their choice of
which way to orient the scooter at each step, and the method by
which the scooter finds the local minimum along the ray determined
by the scooter's orientation.  Finding a minimum along the length
of the ray is the problem of finding the minimum of a function of
one variable. Routines for doing this are described in several
popular texts\cite{linmin}.  In this paper we focus on the many
variables aspect of the problem, the choice of which way to point
the scooter, and in particular those choices that correspond to
conjugate gradient and variable metric minimization.

Conjugate gradient and variable metric (sometimes called
quasi-newton) optimization methods are very popular and successful
algorithms for finding minima of functions of many
variables\cite{cgvm}. In this paper we motivate these optimization
methods, and derive them in a unified manner in a way that
emphasizes their essential similarity. The derivation in this
paper encompasses the conjugate gradient method with and without
conditioning, and the variable metric method. The common variants
of these methods including Fletcher-Reeves\cite{fr},
Polak-Ribiere\cite{pr}, Davidon-Fletcher-Powell\cite{dfp}, and
Broyden-Fletcher-Goldfarb-Shanno\cite{bfgs}  are described and
motivated. In section II we describe what is wrong with the
steepest descent method.  In section III we derive a generalized
optimization routine which is the starting point for all of the
algorithms mentioned above. In section IV we take up some
convergence proofs which are followed up in the Appendices. In
section V we focus on conjugate gradient routines with and without
conditioning and in section VI we turn to variable metric methods.

\section{What's wrong with steepest descent?}
In this section we introduce numerical minimization of a function
of many variables and show why the steepest descent minimization
method is not an optimal choice.  Consider a function $E({\bf x})$
of m unconstrained real variables collectively denoted by the
m-component vector ${\bf x}$. For concreteness you can consider
the function to be the energy of a configuration of ions. We seek
the position of the ions for which the energy is minimized.  As
with finding the lowest point in a valley, we can think in terms
of orienting and moving a single magic scooter but now the scooter
moves in an m-dimensional space in which each point corresponds to
a configuration of all of the ions. For example, if there are two
ions labeled by coordinates $(x_1,y_1,z_1)$ and $(x_2,y_2,z_2)$
the scooter moves in the six dimensional space ${\bf
x}=(x_1,y_1,z_1,x_2,y_2,z_2)$ .  In general the n+1 guess for the
location of the scooter is determined by
\begin{equation}
{\bf x}_{n+1}={\bf x}_n+\lambda_n {\bf h}_n \label{xn+1}
\end{equation}
where ${\bf h}_n$ is a vector, different for each minimization
scheme, that points along the direction the scooter is headed and
$\lambda_n$ is a positive scaler that represents how far the
scooter moves, in units of $|{\bf h}_n|$. Once ${\bf h}_n$ is
chosen we find $x_{n+1}$ by varying $\lambda_n$ to minimize
$E({\bf x}_{n+1})$. The minimum is found where the derivative of
the energy $E$ with respect to $\lambda_n$ is zero,
\begin{equation}
\frac{\partial E}{\partial \lambda_n} = {\bf \nabla} E({\bf
x}_{n+1})\cdot {\bf h}_n =0,
\end{equation}
or by multiplying by $\lambda_n$
\begin{equation}
{\bf \nabla} E({\bf x}_{n+1})\cdot ({\bf x}_{n+1}-{\bf x}_n)
=0\label{alwaystrue}.
\end{equation}
This result tells us that the gradient, ${\bf \nabla} E({\bf
x}_{n+1})$, at the new minimum ${\bf x}_{n+1}$ is perpendicular to
the descent direction, ${\bf x}_{n+1}-{\bf x}_n$, that took us to
the minimum.  This result is easy to visualize. The dot-product of
the gradient of the energy with a particular vector represents the
change in energy along that vector.  The scooter stops when this
is zero since that indicates that the energy is no longer
decreasing in that direction.

Numerical minimization schemes differ in their choice for ${\bf
h}_n$. A naive choice is the steepest decent direction at ${\bf
x}_n$~,
\begin{equation}
{\bf h}_n=-{\bf \nabla} E({\bf x}_n).
\end{equation}
As well described in standard texts\cite{cgvm}, this choice turns
out to be far from optimal since it forces each new decent
direction to be orthogonal to the previous one. The path to the
minimum using this method is a series of 90 degree jogs and it can
take many steps of this sort to get close to the true minimum.
Consider, for example, a valley with a long shallow axis and a
short steep axis shaped like a canoe or a long wooden stirring
spoon. Start the scooter somewhere away from the long axis.  The
steepest descent direction will point somewhere between the
shortest path to the long axis, and the direct path to the
minimum.  Most people assume that the scooter will stop at the
long axis and the next move will be along the long axis to the
minimum. But the lowest point along the ray must overshoot the
long axis by some amount for the next steepest descent direction
to be 90 degrees from the path just taken. This is because the
energy lost by going parallel to the long axis isn't overtaken by
the gain in going up the steep side until some distance beyond the
long axis. For this reason, the steepest descent path constantly
overshoots, taking many short 90 degree jogs along the long axis
on the way to the minimum, when only two steps appear to be
necessary. The way out of this dilemma, adopted by the conjugate
gradient and variable metric methods, is to use some other
direction than steepest descent for the descent direction. This is
what we turn to in the next section.

\section{A Universal Optimization Procedure}
In this section we improve upon the steepest descent method and
develop a universal framework for optimization. The basic idea is
to pick a descent direction that has some positive projection onto
the steepest descent direction but also incorporates information
about previous moves into our n+1 guess. A natural choice, as we
will see below, for the n+1 guess constructs the direction vector
from a linear combination of all previous displacements plus a new
vector ${\bf v}_n$ that incorporates new information about $E$ at
${\bf x}_n$,
\begin{equation}
{\bf x}_{n+1} = {\bf x}_n +\lambda_n\left({\bf v}_n
+\sum_{j=2}^na_{j,n}({\bf x}_j-{\bf x}_{j-1})\right). \label{hn}
\end{equation}
For conjugate gradient minimization ${\bf v}_n = -{\bf
\nabla}E(x_n)$. For conjugate gradient with conditioning or
variable metric minimization ${\bf v}_n=-{\bf H}_n{\bf
\nabla}E(x_n)$ where ${\bf H}_n$ is an appropriately chosen
symmetric matrix called the conditioner if it is independent of n
or the variable metric if it depends on n. The variables
$\lambda_n$ and $a_{j,n}$ are varied either analytically or
numerically to minimize the function in question. We will minimize
E analytically with respect to the $a_{j,n}$'s. E is then a
function of the single variable $\lambda_n$ and can be minimized
numerically using a routine to find the minimum of a function of
one variable. Choosing ${\bf v}_n$ and the $a_{j,n}$ beforehand
determines ${\bf h}_n$ and is like orienting the scooter.
Determining $\lambda_n$ numerically is like taking a scooter ride.

When we take the derivative of the energy $E$ with respect to
$\lambda_n$ at the point ${\bf x}_{n+1}$ and set it to zero, we
get Eq.(\ref{alwaystrue}), as before, but now ${\bf \nabla} E({\bf
x}_{n+1})$ is not the next descent direction so this condition
does not constrain us to right angle turns as was the case for
steepest descent minimization.

We get n-1 new conditions when we take the derivative of the
energy $E$ with respect to $a_{j,n}$  at the point ${\bf x}_{n+1}$
and set it to zero (In practice it is always clear that we are
dealing with the n+1 guess so we have throughout the rest of this
paper suppressed the second index in the $a_{j,n}$.),
\begin{equation}
\frac{\partial E}{\partial a_j} = \lambda_n{\bf \nabla} E({\bf
x}_{n+1})\cdot ({\bf x}_j-{\bf x}_{j-1})=0\;\;\; 2\le j\le
n.\label{dedaj}
\end{equation}

Divided by $\lambda_n$ and combined with Eq.(\ref{alwaystrue}) we
have
\begin{equation}
{\bf \nabla} E({\bf x}_{n+1})\cdot ({\bf x}_j-{\bf
x}_{j-1})=0\;\;\; 2\le j\le n+1.\label{orthog}
\end{equation}
which tells us that the gradient at step $n+1$ is orthogonal to
all previous displacements.  These relations will be a great aid
in simplifying expressions.

We will get back to Eq.(\ref{dedaj}) shortly but first we need to
express ${\bf \nabla} E({\bf x}_{n+1})$ in terms of the $a_j$'s
and known quantities. To do so we assume an explicit form for
$E({\bf x})$. We develop our optimization scheme assuming that our
guess for the minimum is close enough to the true minimum ${\bf
x}_{\rm min}$ that we can approximate $E({\bf x})$ by its Taylor
expansion to second order around its true minimum.
\begin{equation}
E({\bf x}) = E({\bf x}_{\rm min}) + \frac{1}{2}({\bf x}-{\bf
x}_{\rm min})\cdot {\bf A} \cdot({\bf x}-{\bf x}_{\rm min}),
\label{E}
\end{equation}
where the first order term is zero since we are expanding $E({\bf
x})$ around its minimum. The $m\times m$ symmetric constant matrix
${\bf A}$ is the diadic second derivative of $E({\bf x})$
evaluated at ${\bf x}_{\rm min}$ with matrix elements
$A_{i,j}=\partial^2E({\bf x}_{\rm min})/\partial x_i\partial x_j$.
For a unique minimum, ${\bf A}$ must have only positive
eigenvalues.

The quadratic form for the energy Eq.(\ref{E}) allows us to find
an explicit expression for the gradient
\begin{equation}
{\bf \nabla} E({\bf x}) = {\bf A}\cdot({\bf x}-{\bf x}_{\rm min})
= ({\bf x}-{\bf x}_{\rm min})\cdot {\bf A}\label{xbar}.
\end{equation}
However, for this to be useful we must express it in such a way
that ${\bf x}_{\rm min}$ and ${\bf A}$ do not explicitly appear
since they are unknown. We can get rid of ${\bf x}_{\rm min}$ by
subtracting the gradients at any two points in configuration
space,
\begin{equation}
{\bf \nabla} E({\bf x}) -{\bf \nabla} E({\bf y}) = {\bf
A}\cdot({\bf x}-{\bf y})=({\bf x}-{\bf y})\cdot {\bf A}.
\label{Arule}
\end{equation}
This is an incredibly useful formula since it allows us to turn
any expression with {\bf A}, which we don't know, into quantities
we do know as long as we can arrange for {\bf A} to operate on a
displacement. This is the reason why including previous
displacements in our n+1 guess turns out to be the natural way to
incorporate information from previous moves.  We now have what we
need to manipulate Eq.(\ref{dedaj}) to find the n-1 $a_j$'s.

First we use Eq.(\ref{Arule}) to find an explicit expression for
${\bf \nabla} E({\bf x}_{n+1})$ in Eq.({\ref{dedaj}) by setting
${\bf x}= {\bf x}_{n+1}$ and ${\bf y}={\bf x}_n$,
\begin{equation}
{\bf \nabla} E({\bf x}_{n+1}) = {\bf \nabla} E({\bf x}_{n}) +
({\bf x}_{n+1}-{\bf x}_n)\cdot {\bf A}.
\end{equation}

We now use this expression for ${\bf \nabla} E({\bf x}_{n+1})$ in
Eq.(\ref{dedaj}) to get, after dividing by $\lambda_n$,
\begin{equation}
{\bf \nabla} E({\bf x}_{n})\cdot ({\bf x}_j-{\bf x}_{j-1}) + ({\bf
x}_{n+1}-{\bf x}_n)\cdot {\bf A} \cdot ({\bf x}_j-{\bf x}_{j-1})=0
\;\;\; 2\le j\le n\label{dedaj2}.
\end{equation}
The first term in Eq.(\ref{dedaj2}) is zero via Eq.(\ref{orthog})
with n+1 replaced with n. We are left with
\begin{equation}
({\bf x}_{n+1}-{\bf x}_n)\cdot {\bf A} \cdot ({\bf x}_j-{\bf
x}_{j-1})=0 \;\;\;  2\le j\le n \label{dedaj3}.
\end{equation}

We can now, as promised, get rid of the unknown ${\bf A}$ via
Eq.({\ref{Arule}) since it is operating on a displacement.
\begin{equation}
({\bf x}_{n+1}-{\bf x}_n)\cdot \left({\bf \nabla} E({\bf x}_j)
-{\bf \nabla} E({\bf x}_{j-1})\right)=0 \;\;\; 2\le j\le n
\label{dedaj3.5}.
\end{equation}

When we put the formula for ${\bf x}_{n+1}$, Eq.(\ref{hn}), into
Eqs.(\ref{dedaj3.5}) we get, after dividing by $\lambda_n$, what
we have been after since the beginning of this section: n-1
equations for the n-1 $a_j$'s in terms of known quantities from
previous steps,
\begin{equation}
\left({\bf v}_n + \sum_{k=2}^na_k({\bf x}_k-{\bf
x}_{k-1})\right)\cdot\left({\bf \nabla} E({\bf x}_j) -{\bf \nabla}
E({\bf x}_{j-1})\right) =0\;\;\; 2\le j\le n\label{dedaj4}.
\end{equation}
By inverting these equations we can find a direction to orient our
scooter that is closer to optimal the closer we are to the minimum
and the closer the quadratic approximation is to $E$. This is
already more than we had any right to expect but it gets even
better! These n-1 equations are uncoupled since by combining
Eq.(\ref{Arule}) with Eq.(\ref{dedaj3}) and using the fact that
{\bf A} is symmetric, we have
\begin{eqnarray}
({\bf x}_{k}-{\bf x}_{k-1})\cdot {\bf A} \cdot ({\bf x}_j-{\bf
x}_{j-1}) &=& ({\bf x}_{k}-{\bf x}_{k-1})\cdot\left({\bf \nabla}
E({\bf x}_j) -{\bf \nabla} E({\bf
x}_{j-1})\right)\nonumber\\&=&\left({\bf \nabla} E({\bf x}_k)
-{\bf \nabla} E({\bf x}_{k-1})\right)\cdot ({\bf x}_j-{\bf
x}_{j-1})\nonumber\\&=&\delta_{j,k}\left({\bf \nabla} E({\bf x}_j)
-{\bf \nabla} E({\bf x}_{j-1})\right)\cdot ({\bf x}_j-{\bf
x}_{j-1})\label{bigdeal}
\end{eqnarray}
so that all terms in the k-sum of Eq.(\ref{dedaj4}) drop out
except the k=j term.  The resulting formula for $a_j$ is
\begin{equation}
a_j = -\frac{{\bf v}_n\cdot({\bf \nabla} E({\bf x}_j) -{\bf
\nabla} E({\bf x}_{j-1}))}{({\bf \nabla} E({\bf x}_j) -{\bf
\nabla} E({\bf x}_{j-1}))\cdot ({\bf x}_j-{\bf x}_{j-1})}.
\end{equation}
Plugging this into the update formula Eq.(\ref{hn}) we obtain the
expression that all of the optimization schemes take as a starting
point.
\begin{equation}
{\bf x}_{n+1}={\bf x}_n + \lambda_n\left({\bf v}_n -
\sum_{j=1}^n({\bf x}_j-{\bf x}_{j-1})\frac{{\bf v}_n\cdot({\bf
\nabla} E({\bf x}_j) -{\bf \nabla} E({\bf x}_{j-1}))}{({\bf
\nabla} E({\bf x}_j) -{\bf \nabla} E({\bf x}_{j-1}))\cdot ({\bf
x}_j-{\bf x}_{j-1})}\right)\label{canonical}
\end{equation}
This is the most important equation in this paper.  Conjugate
gradient and variable metric routines use this formula and only
differ in their choice of ${\bf v}_n$.

\section{convergence}
Of course, none of this is worth it unless Eq.(\ref{canonical}) is
demonstrably faster than steepest descent.  And it is, at least
for those cases where one is close enough to a minimum that the
second order Taylor expansion for $E$ is a good approximation. For
quadratic energies, this updating scheme will converge to the
minimum of $E$ in no more steps than the number of components of
${\bf x}$ (two for the canoe, not including the first guess),
provided that the ${\bf v}_n$'s are chosen to be linearly
independent of each other. The proof is straightforward. Consider
the $m+1$ step of Eq.(\ref{canonical}).  As long as ${\bf v}_{m}$
and the $m-1$ $({\bf x}_j-{\bf x}_{j-1})$'s are linearly
independent, which will be true if all the ${\bf v}_n$'s are
linearly independent, then by varying $\lambda_m$ and the $m-1$
$a_j$'s, ${\bf x}_{m+1}$ spans the entire configuration space.
Since $\lambda_m$ and all the $a_j$'s are minimized exactly
(remember the $a_j$'s are exact only for quadratic functions)
${\bf x}_{m+1}$ is guaranteed to be the minimum. So, our intuition
that it should only take two steps to find the minimum of a canoe
is correct, as long as we have a quadratic canoe.

We can achieve the minimum in fewer than $m$ steps with a clever
choice of ${\bf v}_n$, provided that E is exactly given by
Eq.(\ref{E}). For example, the steepest descent method, ${\bf v}_1
= -{\bf \nabla} E({\bf x}_1)$, finds the minimum in one step
regardless of the dimensionality of the configuration space for
the maximally symmetric case of quadratic functions with ${\bf A}$
proportional to the identity matrix (a circular parabolic bowl is
an example). More generally one step minimization for quadratic
functions is always possible with
\begin{equation}
{\bf x}_{\rm min} = {\bf x}_n - {\bf A}^{-1}\cdot{\bf \nabla}
E({\bf x}_n)\label{exact1step},
\end{equation}
found by inverting Eq.(\ref{xbar}). This motivates choosing ${\bf
v}_n = -{\bf H}_n\cdot{\bf \nabla} E({\bf x}_n)$ where the matrix
${\bf H}_n$ is a possibly different symmetric positive definite
matrix at each step and is ideally a good approximation for ${\bf
A}^{-1}$. Different choices for ${\bf H}_n$ correspond to
different conjugate gradient and variable metric routines.

In the appendix we show that optimization of quadratic functions
with appropriate choices for ${\bf H}_n$ converges to the minimum
of $E$ in no more steps than the number of distinct eigenvalues of
${\bf H}_1{\bf A}$, which can be much less than m for highly
symmetric systems (notice that this agrees with 1 step
minimization when ${\bf H}_1$ is proportional to ${\bf A}^{-1}$).
In the Appendix we also show that these schemes automatically
construct a better approximation for ${\bf A}^{-1}$ at each step
so the optimal choice for $\lambda_n$ approaches $1$ for large
$n$, regardless of how poorly ${\bf H}_1$ initially approximates
${\bf A}^{-1}$.

Even if the energy is not well approximated by a quadratic
function, positive definite ${\bf H}_n$ still guarantees that the
positive $\lambda_n$ direction points down hill.  The proof is
straightforward. Downhill is determined by
\begin{equation}
{\bf \nabla} E({\bf x}_n)\cdot ({\bf x}_{n+1}- {\bf x}_n) <0.
\end{equation}
Replacing the displacement ${\bf x}_{n+1}- {\bf x}_n$ with
Eq.(\ref{canonical}) and using Eq.(\ref{orthog}) to remove all
terms in the $j$ sum we have
\begin{equation}
-\lambda_n {\bf \nabla} E({\bf x}_n)\cdot {\bf H}_n{\bf \nabla}
E({\bf x}_n)<0,
\end{equation}
which guarantees that $E$ decreases in the positive $\lambda_n$
direction, at least initially, provided ${\bf H}_n$ is positive
definite.

If the energy function is not quadratic, convergence to the
minimum is not guaranteed for any finite number of steps since the
$a_j$'s given by this procedure are no longer the true minima.
Regardless, these methods are generally better than steepest
descent, since smooth functions are generally closer to quadratic
the closer one is to the minimum so the approximate $a_j$'s get
progressively closer to their true values.  Many implementations
of these methods switch over to steepest descent minimization in
the event that ${\bf x}_n$ is in a region that is sufficiently far
from quadratic.

\section{Conjugate Gradient Optimization}
We are now ready to derive conjugate gradient optimization
formulas. This section will be short since we have already done
most of the work we need to do. Conjugate gradient optimization is
just the general optimization procedure, Eq.(\ref{canonical}) with
${\bf v}_n = -{\bf H}\cdot{\bf \nabla} E({\bf x}_n)$ where ${\bf
H}$, called the conditioner, is an appropriately chosen symmetric
positive definite matrix. This is motivated by the exact result in
case of quadratic functions Eq.({\ref{exact1step}).  If one has a
good estimate for ${\bf A}^{-1}$ then one can use it for ${\bf
H}$. Often no such information is known and the identity matrix is
used for {\bf H}. The updating formula is
\begin{equation} {\bf x}_{n+1} = {\bf x}_{n} -\lambda_n \left(
{\bf H}\cdot{\bf \nabla} E({\bf x}_n) - \sum_{j=2}^n({\bf
x}_j-{\bf x}_{j-1})\frac{({\bf \nabla} E({\bf x}_j) -{\bf \nabla}
E({\bf x}_{j-1}))\cdot {\bf H} \cdot{\bf \nabla} E({\bf
x}_n)}{({\bf x}_j-{\bf x}_{j-1})\cdot({\bf \nabla} E({\bf
x}_{j})-{\bf \nabla} E({\bf x}_{j-1})) }\right).\label{cg1}
\end{equation}

The conjugate gradient choice for ${\bf v}_n$ has the wonderful
feature that it makes all terms in the $j$ sum in Eq.(\ref{cg1})
equal to zero except for the j=n term.  To see this note that

\begin{equation}
{\bf \nabla} E({\bf x}_n)\cdot {\bf H} \cdot{\bf \nabla} E({\bf
x}_j)={\bf \nabla} E({\bf x}_n)\cdot \left ({\bf H} \cdot {\bf
\nabla} E({\bf x}_j) - \sum_{k=2}^j({\bf x}_k-{\bf
x}_{k-1})\frac{({\bf \nabla} E({\bf x}_k) -{\bf \nabla} E({\bf
x}_{k-1}))\cdot {\bf H} \cdot{\bf \nabla} E({\bf x}_j)}{({\bf
x}_k-{\bf x}_{k-1})\cdot({\bf \nabla} E({\bf x}_{k})-{\bf \nabla}
E({\bf x}_{k-1})) }\right),
\end{equation}
for $j\le n$, where we have added terms that are zero via
Eq.(\ref{orthog}). Now using Eq.(\ref{cg1}) we can replace the
terms in parenthesis with the displacement so we get
\begin{equation}
{\bf \nabla} E({\bf x}_{n})\cdot {\bf H}\cdot {\bf \nabla}E({\bf
x}_j) = {\bf \nabla} E({\bf x}_{n})\cdot ({\bf x}_{j+1}-{\bf
x}_j)/\lambda_j,
\end{equation}
which by Eq.(\ref{orthog}) is zero for $j <n-1$, so that we have
the result
\begin{equation}
{\bf \nabla} E({\bf x}_{n})\cdot {\bf H}\cdot {\bf \nabla}E({\bf
x}_j) = 0\;\;\; 2\le j\le n-1 \label{good}
\end{equation}

Eq.(\ref{good}) makes all terms in the $j$ sum zero in
Eq.(\ref{cg1}) except for $j=n$.  We are left with
\begin{equation}
{\bf x}_{n+1} = {\bf x}_{n} -\lambda_n \left( {\bf H}\cdot{\bf
\nabla} E({\bf x}_n) - ({\bf x}_n-{\bf x}_{n-1})\frac{({\bf
\nabla} E({\bf x}_n) -{\bf \nabla} E({\bf x}_{n-1}))\cdot {\bf H}
\cdot{\bf \nabla} E({\bf x}_n)}{({\bf x}_n-{\bf
x}_{n-1})\cdot({\bf \nabla} E({\bf x}_{n})-{\bf \nabla} E({\bf
x}_{n-1})) }\right).\label{PR}
\end{equation}
This expression for ${\bf x}_{n+1}$ is the Polak-Ribiere conjugate
gradient updating formula (However, see Appendix A). One can come
up with dozens of other formulas simply by adding or subtracting
terms that are zero if the function is quadratic. Another popular
choice is the Fletcher-Reeves updating formula which, using the
fact that ${\bf \nabla} E({\bf x}_{n-1})\cdot {\bf H}\cdot{\bf
\nabla} E({\bf x}_n) =0$ from Eq.(\ref{good}), simplifies
Eq.(\ref{PR}) to
\begin{equation}
{\bf x}_{n+1} = {\bf x}_{n} -\lambda_n \left( {\bf H}\cdot{\bf
\nabla} E({\bf x}_n) - ({\bf x}_n-{\bf x}_{n-1})\frac{{\bf \nabla}
E({\bf x}_n)\cdot {\bf H} \cdot{\bf \nabla} E({\bf x}_n)}{({\bf
x}_n-{\bf x}_{n-1})\cdot({\bf \nabla} E({\bf x}_{n})-{\bf \nabla}
E({\bf x}_{n-1})). }\right)\label{FR}
\end{equation}
The Polak-Ribiere and the Fletcher-Reeves formulas are equal if
the energy function is quadratic. If the energy function is not
quadratic it is possible for the Polak-Ribiere numerator to be
negative, while the Fletcher-Reeves numerator is always positive.
Empirically the best form for the numerator appears to be ${\rm
max}\left(({\bf \nabla} E({\bf x}_n) -{\bf \nabla} E({\bf
x}_{n-1}))\cdot {\bf H} \cdot {\bf \nabla} E({\bf x}_n),0\right)$
which uses the Polak-Ribiere numerator but restarts the conjugate
gradient routine with a steepest descent step (assuming the
conditioner is the identity) if the P-R numerator goes negative.

The conjugate gradient update formulas are swell but they don't
exactly look much like the one step formula Eq.(\ref{exact1step})
that we might have expected them to resemble. This can be remedied
by taking advantage of the ability to create matrices by forming
the outer product of two vectors. For the Polak-Ribiere formula we
can write
\begin{equation}
{\bf x}_{n+1} = {\bf x}_{n} -\lambda_n \left( {\bf H} -
\frac{({\bf x}_n-{\bf x}_{n-1})\bigotimes({\bf \nabla} E({\bf
x}_n) -{\bf \nabla} E({\bf x}_{n-1}))\cdot {\bf H} }{({\bf
x}_n-{\bf x}_{n-1})\cdot({\bf \nabla} E({\bf x}_{n})-{\bf \nabla}
E({\bf x}_{n-1})) }\right)\cdot{\bf \nabla} E({\bf
x}_n)\label{pullout},
\end{equation}
in which $\bigotimes$ connotes the outer product. A similar
expression holds for the Fletcher-Reeves update formula. This
change to the form of the descent direction is entirely cosmetic,
and is given primarily to show the similarity with
Eq.(\ref{exact1step}). However, Eq.(\ref{pullout} suggests the
possibility that the expression in parenthesis, modified to
exhibit symmetry, positive definiteness and other properties we
expect of an approximation for ${\bf A}^{-1}$, would be an
improvement over ${\bf H}$ that could profitably be used in the
next update in its place. This is the motivation for variable
metric minimization which is discussed in the next section.

\section{variable metric optimization}
For variable metric optimization we choose ${\bf v}_n = -{\bf
H}_n\cdot{\bf \nabla} E({\bf x}_n)$ where ${\bf H}_n$ is an
appropriately chosen symmetric positive definite matrix that is
updated at each step to more closely resemble ${\bf A}^{-1}$. The
initial ${\bf H}_1$ is often the identity matrix. ${\bf A}^{-1}$
has the useful property determined by the inverse of
Eq.(\ref{Arule}) that
\begin{equation}
{\bf x}-{\bf y}= {\bf A}^{-1}\cdot({\bf \nabla} E({\bf x}) -{\bf
\nabla} E({\bf y})) = ({\bf \nabla} E({\bf x}) -{\bf \nabla}
E({\bf y}))\cdot {\bf A}^{-1}\label{a-1},
\end{equation}
which is good for any ${\bf x}$ and ${\bf y}$.  Since ${\bf H}_n$
is supposed to be a good approximation for ${\bf A}^{-1}$ we will
require that ${\bf H}_n$ is symmetric and satisfies Eq.(\ref{a-1})
for ${\bf x}$ and ${\bf y}$ chosen from the set of ${\bf x}_j$ for
$j\le n$.  This will be so if we require
\begin{equation}
{\bf H}_n\cdot({\bf \nabla} E({\bf x}_j) -{\bf \nabla} E({\bf
x}_{j-1}))={\bf x}_j - {\bf x}_{j-1}\;\;\;2\le j\le
n.\label{hrule}
\end{equation}
For an ${\bf H}_n$ satisfying Eq.(\ref{hrule}) the update formula
\begin{equation}
{\bf x}_{n+1} = {\bf x}_{n} -\lambda \left( {\bf H}_n\cdot{\bf
\nabla} E({\bf x}_n) - \sum_{j=2}^n({\bf x}_j-{\bf
x}_{j-1})\frac{({\bf \nabla} E({\bf x}_j) -{\bf \nabla} E({\bf
x}_{j-1}))\cdot {\bf H}_n \cdot {\bf \nabla} E({\bf x}_n)}{({\bf
\nabla} E({\bf x}_j) -{\bf \nabla} E({\bf x}_{j-1}))\cdot ({\bf
x}_j-{\bf x}_{j-1})}\right)\label{Hcanonical}
\end{equation}
simplifies to
\begin{equation}
{\bf x}_{n+1} = {\bf x}_{n} -\lambda_n {\bf H}_n\cdot{\bf \nabla}
E({\bf x}_n)\label{qnd}
\end{equation}
via Eqn.'s (\ref{hrule}) and (\ref{orthog}).  The resulting
expression is naturally in the form resembling
Eq.(\ref{exact1step}) so ${\bf H}_n$ is acting like an approximate
${\bf A}^{-1}$.

There are many ways to construct an ${\bf H}_n$ that satisfies
Eq.(\ref{hrule}).  An interesting formula results if we choose
${\bf v}_n = -{\bf H}_{n-1}\cdot{\bf \nabla} E({\bf x}_n)$ but
insist that we still end up with Eq.(\ref{qnd}) for ${\bf
x}_{n+1}$. For this to be true we must have
\begin{equation}
{\bf H}_n\cdot{\bf \nabla} E({\bf x}_n) ={\bf H}_{n-1}\cdot{\bf
\nabla} E({\bf x}_n) - ({\bf x}_n-{\bf x}_{n-1})\frac{({\bf
\nabla} E({\bf x}_n) -{\bf \nabla} E({\bf x}_{n-1}))\cdot {\bf
H}_{n-1} \cdot {\bf \nabla} E({\bf x}_n)}{({\bf \nabla} E({\bf
x}_n) -{\bf \nabla} E({\bf x}_{n-1}))\cdot ({\bf x}_n-{\bf
x}_{n-1})}\label{hn-1}
\end{equation}

An ${\bf H}_n$ that satisfies this is
\begin{equation}
{\bf H}_n ={\bf H}_{n-1} - \frac{({\bf x}_n-{\bf
x}_{n-1})\bigotimes {\bf H}_{n-1}\cdot({\bf \nabla} E({\bf x}_n)
-{\bf \nabla} E({\bf x}_{n-1}))}{({\bf \nabla} E({\bf x}_n) -{\bf
\nabla} E({\bf x}_{n-1}))\cdot ({\bf x}_n-{\bf x}_{n-1})},
\end{equation}
which is an iterative form of Eq.(\ref{pullout}).  However, this
${\bf H}_n$ is a bad approximation for ${\bf A}^{-1}$ since it is
neither symmetric nor does it satisfy Eq.(\ref{hrule}). This can
be remedied by amending ${\bf H}_n$ with terms that give zero for
any E, quadratic or not, when operating on ${\bf \nabla} E({\bf
x}_n)$;
\begin{eqnarray}
{\bf H}_n &=&{\bf H}_{n-1} - \frac{({\bf x}_n-{\bf
x}_{n-1})\bigotimes {\bf H}_{n-1}\cdot({\bf \nabla} E({\bf x}_n)
-{\bf \nabla} E({\bf x}_{n-1}))}{({\bf \nabla} E({\bf x}_n) -{\bf
\nabla} E({\bf x}_{n-1}))\cdot ({\bf x}_n-{\bf
x}_{n-1})}-\frac{{\bf H}_{n-1}\cdot({\bf \nabla} E({\bf x}_n)
-{\bf \nabla} E({\bf x}_{n-1}))\bigotimes({\bf x}_n-{\bf x}_{n-1})
}{({\bf \nabla} E({\bf x}_n) -{\bf \nabla} E({\bf x}_{n-1}))\cdot
({\bf x}_n-{\bf x}_{n-1})}\nonumber\\ &+& a({\bf x}_n-{\bf
x}_{n-1})\bigotimes({\bf x}_n-{\bf x}_{n-1})\label{bsc}.
\end{eqnarray}
The next to last term was added to make ${\bf H}_n$ symmetric. The
last term is allowed with any value of $a$.  Remarkably, there is
a special value:
\begin{equation}
a= \frac{1}{({\bf \nabla} E({\bf x}_n) -{\bf \nabla} E({\bf
x}_{n-1}))\cdot ({\bf x}_n-{\bf x}_{n-1})}\left(1 + \frac{({\bf
\nabla} E({\bf x}_n) -{\bf \nabla} E({\bf x}_{n-1}))\cdot {\bf
H}_{n-1}\cdot({\bf \nabla} E({\bf x}_n) -{\bf \nabla} E({\bf
x}_{n-1}))}{({\bf \nabla} E({\bf x}_n) -{\bf \nabla} E({\bf
x}_{n-1}))\cdot ({\bf x}_n-{\bf x}_{n-1})}\right)\label{besta}
\end{equation}
for which Eq.(\ref{hrule}) is automatically satisfied, so that we
are improving our approximation for ${\bf A}^{-1}$ at every step.
With this value for $a$ Eq.(\ref{bsc}) is the
Broyden-Fletcher-Goldfarb-Shanno variable metric updating formula.
The proof that ${\bf H}_n$ is positive definite, needed to
guarantee that the positive $\lambda_n$ direction points downhill,
is left for the Appendix.

Another popular variable metric update formula, the
Davidon-Fletcher-Powell scheme, was developed with only
Eq.(\ref{hrule}) and symmetry in mind, with no regard for
Eq.(\ref{hn-1}) (although it turns out to satisfy it as a
proportionality rather than an equality as we show in the
Appendix). It is
\begin{eqnarray}
{\bf H}_n &=&{\bf H}_{n-1} - \frac{{\bf H}_{n-1}\cdot({\bf \nabla}
E({\bf x}_n) -{\bf \nabla} E({\bf x}_{n-1}))\bigotimes {\bf
H}_{n-1}\cdot({\bf \nabla} E({\bf x}_n) -{\bf \nabla} E({\bf
x}_{n-1}))}{({\bf \nabla} E({\bf x}_n) -{\bf \nabla} E({\bf
x}_{n-1}))\cdot {\bf H}_{n-1}\cdot({\bf \nabla} E({\bf x}_n) -{\bf
\nabla} E({\bf x}_{n-1}))}\nonumber\\ &+& \frac{({\bf x}_n-{\bf
x}_{n-1})\bigotimes({\bf x}_n-{\bf x}_{n-1})}{({\bf \nabla} E({\bf
x}_n) -{\bf \nabla} E({\bf x}_{n-1}))\cdot ({\bf x}_n-{\bf
x}_{n-1})}\label{dfp}.
\end{eqnarray}
This is an arbitrary formula in that we can add the term $b {\bf
u}\bigotimes {\bf u}$ to ${\bf H}_n$ with
\begin{equation}
{\bf u} = \frac{({\bf x}_n-{\bf x}_{n-1})}{({\bf \nabla} E({\bf
x}_n) -{\bf \nabla} E({\bf x}_{n-1}))\cdot ({\bf x}_n-{\bf
x}_{n-1})} - \frac{{\bf H}_{n-1}\cdot({\bf \nabla} E({\bf x}_n)
-{\bf \nabla} E({\bf x}_{n-1}))}{({\bf \nabla} E({\bf x}_n) -{\bf
\nabla} E({\bf x}_{n-1}))\cdot {\bf H}_{n-1}\cdot({\bf \nabla}
E({\bf x}_n) -{\bf \nabla} E({\bf x}_{n-1}))}
\end{equation}
for any $b$ and still obey Eq.(\ref{hrule}).  The choice
\begin{equation}
b=({\bf \nabla} E({\bf x}_n) -{\bf \nabla} E({\bf x}_{n-1}))\cdot
{\bf H}_{n-1}\cdot({\bf \nabla} E({\bf x}_n) -{\bf \nabla} E({\bf
x}_{n-1}))
\end{equation}
corresponds to the BFGS scheme. It also is the only choice that
makes ${\bf H}_n$ linear in ${\bf H}_{n-1}$. The
Davidon-Fletcher-Powell scheme, for any b, is exactly equivalent
to the Broyden-Fletcher-Goldfarb-Shanno scheme if the energy
function is quadratic. Empirically, for more general nonquadratic
energies, the Broyden-Fletcher-Goldfarb-Shanno scheme is usually
better than the $b=0$ Davidon-Fletcher-Powell scheme.

For quadratic energy functions variable metric and conjugate
gradient methods are equivalent. The proof is in the appendix c.
Variable metric minimization differs from conjugate gradient
minimization for non-quadratic energy functions.

The variable metric method requires more memory than the conjugate
gradient method since for variable metric optimization one is
building up an $m\times m$ matrix.  In practice, for large m, one
can cut this way down by keeping information from only the last
$q<m$ steps\cite{sparce}. My experience has been that for
nonquadratic functions the variable metric method with $q>1$
converges faster than the conjugate gradient method (equivalent to
variable metric with $q=1$) but past performance is no guarantee
of future returns.

\appendix
\section{Will the real conjugate gradient formula please stand up?}
Eq.(\ref{PR}) is the Polak-Ribiere conjugate gradient update
formula and Eq.(\ref{FR}) is the Fletcher-Reeves formula. However,
the conventions for conjugate gradient minimization and variable
metric minimization are different. In this paper we have followed
the variable metric convention throughout so our expressions for
conjugate gradient updates are disguised. In what follows we make
several changes in the appearance of Eq.(\ref{PR}) and
Eq.(\ref{FR}) relying only on the orthogonality of a gradient with
the most recent displacement, Eq.(\ref{alwaystrue}), which is
always true regardless if the energy function is quadratic or not.
The resulting formulas are the conventional expressions for the cg
formulas that appear in most text books.

First we can simplify the denominator in the Polak-Ribiere
routine.
\begin{equation}
{\bf x}_{n+1} = {\bf x}_{n} -\lambda_n \left( {\bf H}\cdot{\bf
\nabla} E({\bf x}_n) + ({\bf x}_n-{\bf x}_{n-1})\frac{({\bf
\nabla} E({\bf x}_n) -{\bf \nabla} E({\bf x}_{n-1}))\cdot {\bf H}
\cdot{\bf \nabla} E({\bf x}_n)}{({\bf x}_n-{\bf x}_{n-1})\cdot{\bf
\nabla} E({\bf x}_{n-1}) }\right)
\end{equation}

One is also always free, from the definition Eq.(\ref{xn+1}), to
express the update scheme in terms of ${\bf h}_{n-1}$ instead of
${\bf x}_n -{\bf x}_{n-1}$.
\begin{equation}
{\bf x}_{n+1} = {\bf x}_{n} -\lambda_n \left( {\bf H}\cdot{\bf
\nabla} E({\bf x}_n) + {\bf h}_{n-1}\frac{({\bf \nabla} E({\bf
x}_n) -{\bf \nabla} E({\bf x}_{n-1}))\cdot {\bf H} \cdot{\bf
\nabla} E({\bf x}_n)}{{\bf h}_{n-1}\cdot{\bf \nabla} E({\bf
x}_{n-1}) }\right)
\end{equation}

Next we expand the denominator using the cg form for ${\bf
h}_{n-1}$.
\begin{eqnarray}
{\bf h}_{n-1}\cdot{\bf \nabla} E({\bf x}_{n-1}) &=& -\left({\bf
H}\cdot{\bf \nabla} E({\bf x}_{n-1}) + {\bf h}_{n-2}\frac{({\bf
\nabla} E({\bf x}_{n-1}) -{\bf \nabla} E({\bf x}_{n-2}))\cdot {\bf
H} \cdot{\bf \nabla} E({\bf x}_{n-1})}{{\bf h}_{n-2}\cdot{\bf
\nabla} E({\bf x}_{n-1}) }\right)\cdot{\bf \nabla} E({\bf
x}_{n-1})\nonumber\\ &=& -{\bf \nabla} E({\bf x}_{n-1})\cdot {\bf
H} \cdot {\bf \nabla} E({\bf x}_{n-1}),
\end{eqnarray}
where the second term was zero by orthogonality,
Eq.(\ref{alwaystrue}). The update formula now looks like
\begin{equation}
{\bf x}_{n+1} = {\bf x}_{n} -\lambda_n \left( {\bf H}\cdot{\bf
\nabla} E({\bf x}_n) - {\bf h}_{n-1}\frac{({\bf \nabla} E({\bf
x}_n) -{\bf \nabla} E({\bf x}_{n-1}))\cdot {\bf H} \cdot{\bf
\nabla} E({\bf x}_n)}{{\bf \nabla} E({\bf x}_{n-1})\cdot {\bf
H}\cdot{\bf \nabla} E({\bf x}_{n-1}) }\right)
\end{equation}
This is the formula that is conventionally used in implementations
of the Polak-Ribiere update formula.

Similarly, the conventional Fletcher-Reeves update is
\begin{equation}
{\bf x}_{n+1} = {\bf x}_{n} -\lambda_n \left( {\bf H}\cdot{\bf
\nabla} E({\bf x}_n) - {\bf h}_{n-1}\frac{{\bf \nabla} E({\bf
x}_n)\cdot {\bf H} \cdot{\bf \nabla} E({\bf x}_n)}{{\bf \nabla}
E({\bf x}_{n-1})\cdot {\bf H}\cdot{\bf \nabla} E({\bf x}_{n-1})
}\right),
\end{equation}
which we remind the reader is only equivalent to the Polak-Ribiere
update formula for quadratic energy functions.

\section{proof that ${\bf H}_n$ is positive definite provided ${\bf H}_{n-1}$ is positive definite}

When finding the minimum of a function it is important to know
which way is down.  As shown in section IV, for conjugate gradient
and variable metric minimization, downhill is in the direction of
positive $\lambda_n$ provided that ${\bf H}_n$ is positive
definite.  For conjugate gradient minimization ${\bf H}_n$ does
not change with n so positive definiteness is trivially guaranteed
by making ${\bf H}$ positive definite initially.  For variable
metric minimization, on the other hand ${\bf H}_n$ is updated at
each step so it is not obvious that if ${\bf H}_1$ is chosen to be
positive definite that ${\bf H}_n$ will be too.

We will simultaneously show that the
Broyden-Fletcher-Goldfarb-Shanno and Davidon-Fletcher-Powell forms
of ${\bf H}_n$ are positive definite provided that ${\bf H}_{n-1}$
is positive definite, regardless if the energy is quadratic or
not. We start with the BFGS expression for ${\bf H}_{n}$,
Eq.(\ref{bsc}), and the DFP expression for ${\bf H}_{n}$,
Eq.(\ref{dfp}) written in the alternative form\cite{sparce}
\begin{equation}
{\bf H}_n = (1-{\bf S}^\dagger)\cdot {\bf H}_{n-1} \cdot(1-{\bf
S}) + \frac{({\bf x}_n-{\bf x}_{n-1})\bigotimes({\bf x}_n-{\bf
x}_{n-1})}{({\bf \nabla} E({\bf x}_n) -{\bf \nabla} E({\bf
x}_{n-1}))\cdot ({\bf x}_n-{\bf x}_{n-1})},
\end{equation}
where
\begin{eqnarray}
{\bf S} &=& \frac{({\bf \nabla} E({\bf x}_n) -{\bf \nabla} E({\bf
x}_{n-1}))\bigotimes({\bf x}_n-{\bf x}_{n-1})}{({\bf \nabla}
E({\bf x}_n) -{\bf \nabla} E({\bf x}_{n-1}))\cdot ({\bf x}_n-{\bf
x}_{n-1})}\;\;\; {\rm BFGS}\nonumber\\ &=& \frac{({\bf \nabla}
E({\bf x}_n) -{\bf \nabla} E({\bf x}_{n-1}))\bigotimes({\bf
\nabla} E({\bf x}_n) -{\bf \nabla} E({\bf x}_{n-1})){\bf H}_{n-1}
}{({\bf \nabla} E({\bf x}_n) -{\bf \nabla} E({\bf x}_{n-1}))\cdot
{\bf H}_{n-1}\cdot ({\bf \nabla} E({\bf x}_n) -{\bf \nabla} E({\bf
x}_{n-1}))}\;\;\; {\rm DFP}.
\end{eqnarray}
To prove positive definiteness we show that ${\bf v} \cdot {\bf
H}_n \cdot {\bf v}
>0$ for arbitrary ${\bf v}$.
\begin{equation}
{\bf v} \cdot {\bf H}_n \cdot {\bf v} = {\bf r} \cdot {\bf
H}_{n-1} \cdot {\bf r} + \frac{({\bf v}\cdot({\bf x}_n-{\bf
x}_{n-1}))^2}{({\bf \nabla} E({\bf x}_n) -{\bf \nabla} E({\bf
x}_{n-1}))\cdot ({\bf x}_n-{\bf x}_{n-1})},\label{vm1pos}
\end{equation} in which
\begin{equation}
{\bf r} = (1-{\bf S})\cdot {\bf v}.
\end{equation}
The denominator of the last term is always greater than zero
provided that ${\bf H}_{n-1}$ is positive definite,
\begin{equation}
({\bf \nabla} E({\bf x}_n) -{\bf \nabla} E({\bf x}_{n-1}))\cdot
({\bf x}_n-{\bf x}_{n-1}) = \lambda_{n-1}{\bf \nabla} E({\bf
x}_{n-1})\cdot{\bf H}_{n-1}\cdot{\bf \nabla} E({\bf
x}_{n-1})>0\label{gtzeero}.
\end{equation}
It is also the case that ${\bf r}$ and ${\bf v}\cdot({\bf
x}_n-{\bf x}_{n-1})$ can not both be zero at the same time if
${\bf v}\ne 0$. This is because ${\bf r}=0$ only when ${\bf v}$ is
proportional to $({\bf \nabla} E({\bf x}_n) -{\bf \nabla} E({\bf
x}_{n-1}))$ which makes ${\bf v}\cdot({\bf x}_n-{\bf x}_{n-1})$
unequal to zero via Eq.(\ref{gtzeero}). Therefore, assuming ${\bf
H}_{n-1}$ is positive definite, Eq.(\ref{vm1pos}) is the sum of
two nonnegative numbers, of which at least one must be greater
than zero. Hence ${\bf H}_n$ is positive definite provided that
${\bf H}_{n-1}$ is positive definite. Consequently the positive
$\lambda_n$ direction points down hill for the variable metric
routines just as it does for the conjugate gradient routines.

\section{proof that all of the variable metric and conjugate
gradient schemes are identical for quadratic functions}

In this section we show that all of the variable metric and
conjugate gradient schemes produce the same sequence of ${\bf
x}_n$'s for quadratic functions. We have already shown that the
Polak-Ribiere and the Fletcher Reeves conjugate gradient routines
are equivalent for quadratic functions.  We will now show that the
variable metric routines are equivalent to conjugate gradient
routines, and that therefore they are also equivalent to each
other.  The proof is inductive and relies on the following
proportionality relations for descent directions
\begin{equation}
{\bf H}_n\cdot{\bf \nabla} E({\bf x}_n) \propto {\bf
H}_{k}\cdot{\bf \nabla} E({\bf x}_n) - ({\bf x}_n-{\bf
x}_{n-1})\frac{({\bf \nabla} E({\bf x}_n) -{\bf \nabla} E({\bf
x}_{n-1}))\cdot {\bf H}_{k} \cdot {\bf \nabla} E({\bf x}_n)}{({\bf
\nabla} E({\bf x}_n) -{\bf \nabla} E({\bf x}_{n-1}))\cdot ({\bf
x}_n-{\bf x}_{n-1})}\;\;\; k<n.\label{k}
\end{equation}
The weaker requirement of proportionality rather than equality is
sufficient to guarantee that the descent {\em directions} are the
same for different $k$; the equality of the displacements and
therefore the ${\bf x}_n$'s is recovered when $\lambda_n$ is
varied numerically. This is trivially satisfied for $k=n-1$ in the
Broyden-Fletcher-Goldfarb-Shanno updating formula where the
proportionality is the equality in Eq.(\ref{hn-1}), and was the
motivation for the BFGS update scheme. The proportionality is also
true for $k=n-1$ for the Davidon-Fletcher-Powell updating formula
as we will prove below. Putting off the $k=n-1$ proof for the DFP
formula for now, we will show that for either scheme if
Eq.(\ref{k}) is satisfied for ${\bf H}_k$ then it is also true for
${\bf H}_{k-1}$.  Once this is proven the equivalence between
variable metric and conjugate gradient optimization is achieved by
setting $k=1$. Plugging either the BFGS expression for ${\bf
H}_{k}$, Eq.(\ref{bsc}), or the DFP expression for ${\bf H}_{k}$,
Eq.(\ref{dfp}), into Eq.(\ref{k}) gets us back Eq.(\ref{k}) with
${\bf H}_{k-1}$ provided that $E$ is quadratic (so that we may
make liberal use of Eq.(\ref{orthog})) and that
\begin{equation}
({\bf \nabla} E({\bf x}_k) -{\bf \nabla} E({\bf x}_{k-1}))\cdot
{\bf H}_{k-1} \cdot {\bf \nabla} E({\bf x}_n)=0.\;\;\;
k<n.\label{zrow}
\end{equation}
The second part of this expression is zero via Eq.(\ref{orthog})
since $E({\bf x}_{k-1})\cdot {\bf H}_{k-1}$ is proportional to the
displacement ${\bf x}_k-{\bf x}_{k-1}$.  To show that $E({\bf
x}_{k})\cdot {\bf H}_{k-1}$ is also proportional to a linear
combinations of displacements depends on Eq.(\ref{k}) being true
for $k=n-1$.  If it is then
\begin{equation}
{\bf H}_{k-1}\cdot E({\bf x}_{k}) = ({\bf x}_k-{\bf
x}_{k-1})\frac{({\bf \nabla} E({\bf x}_k) -{\bf \nabla} E({\bf
x}_{k-1}))\cdot {\bf H}_{k-1} \cdot {\bf \nabla} E({\bf
x}_k)}{({\bf \nabla} E({\bf x}_k) -{\bf \nabla} E({\bf
x}_{k-1}))\cdot ({\bf x}_k-{\bf x}_{k-1})} +\gamma
(\bf{x}_{k+1}-\bf{k}),
\end{equation}
where $\gamma$ is a proportionality constant.  This is zero when
dotted into ${\bf \nabla} E({\bf x}_n)$ so the theorem is proved.
Setting $k=1$ in the displacement formula Eq.({\ref{k}) completes
the proof of equivalence of the BFGS variable metric and conjugate
gradient optimization for quadratic functions. To show that DFP
variable metric optimization is also equivalent to conjugate
gradient optimization for quadratic functions we must show that
Eq.(\ref{k}) is satisfied for $k=n-1$ using the DFP ${\bf H}_n$.
To do so we operate with, Eq.(\ref{dfp}), the DFP formula for
${\bf H}_n$ in terms of ${\bf H}_{n-1}$, on the gradient ${\bf
\nabla} E({\bf x}_n)$.  The result is
\begin{equation}
{\bf H}_n\cdot{\bf \nabla} E({\bf x}_n)  ={\bf H}_{n-1}\cdot{\bf
\nabla} E({\bf x}_n) - {\bf H}_{n-1}\cdot({\bf \nabla} E({\bf
x}_n) -{\bf \nabla} E({\bf x}_{n-1}))\frac{ ({\bf \nabla} E({\bf
x}_n) -{\bf \nabla} E({\bf x}_{n-1})){\bf H}_{n-1}\cdot{\bf
\nabla} E({\bf x}_n) }{({\bf \nabla} E({\bf x}_n) -{\bf \nabla}
E({\bf x}_{n-1}))\cdot {\bf H}_{n-1}\cdot({\bf \nabla} E({\bf
x}_n) -{\bf \nabla} E({\bf x}_{n-1}))}
\end{equation}
Combining terms we get
\begin{equation}
{\bf H}_n\cdot{\bf \nabla} E({\bf x}_n) =\frac{{\bf \nabla} E({\bf
x}_{n-1})\cdot{\bf H}_{n-1}\cdot{\bf \nabla} E({\bf
x}_{n-1})}{{\bf \nabla} E({\bf x}_{n})\cdot{\bf H}_{n-1}\cdot{\bf
\nabla} E({\bf x}_{n})} \left({\bf H}_{n-1}\cdot{\bf \nabla}
E({\bf x}_n) - ({\bf x}_n-{\bf x}_{n-1})\frac{({\bf \nabla} E({\bf
x}_n) -{\bf \nabla} E({\bf x}_{n-1}))\cdot {\bf H}_{n-1} \cdot
{\bf \nabla} E({\bf x}_n)}{({\bf \nabla} E({\bf x}_n) -{\bf
\nabla} E({\bf x}_{n-1}))\cdot ({\bf x}_n-{\bf x}_{n-1})}\right),
\end{equation}
where we have used the fact that ${\bf H}_{n-1}\cdot{\bf \nabla}
E({\bf x}_{n-1})$ is proportional to ${\bf x}_{n}-{\bf x}_{n-1}$,
and we have used Eq.(\ref{orthog}) to simplify the proportionality
prefactor.

This completes the proof that all of the standard conjugate
gradient and variable metric optimization routines produce exactly
the same sequence of $x_n$'s for quadratic functions.  We can use
this fact to prove fast convergence of all of these routines for
the case of quadratic functions merely by showing that it is true
for one.  This is what we do in the next section of the appendix.

\section{proof of fast convergence for quadratic functions}
In this section we show that optimization with ${\bf v}_j = - {\bf
H}\cdot {\bf \nabla} E({\bf x}_j)$ converges to the minimum of
quadratic functions in the number of steps equal to the number of
distinct eigenvalues of ${\bf H \cdot A}$, regardless of the
dimension of the configuration space.  Since conjugate gradient
and variable metric minimization is the same for quadratic
functions, it is sufficient to show this for the conjugate
gradient minimization. The proof is based on the one step
minimization formula
\begin{equation}
{\bf x}_{min} = {\bf x} - {\bf A}^{-1}{\bf \nabla} E({\bf
x})\label{onestep},
\end{equation}
and the little known fact, which we prove below, that if a matrix
${\bf M}$ has $p$ distinct eigenvalues then ${\bf M}^{-1}$ is a
polynomial in ${\bf M}$ with highest power $p-1$,
\begin{equation}
{\bf M}^{-1} = \sum_{j=0}^{p-1}c_j({\bf M})^{j}\label{M}.
\end{equation}
With ${\bf M}={\bf HA}$ we have
\begin{equation} {\bf A}^{-1} =
\sum_{j=0}^{p-1}c_{j}({\bf HA})^{j}{\bf H}\label{a-1p}.
\end{equation}
Plugging this into Eq.(\ref{onestep}) we have
\begin{equation}
{\bf x}_{\rm min} = {\bf x} -\sum_{j=0}^{p-1}c_{j}({\bf
HA})^{j}{\bf H}\cdot{\bf \nabla} E({\bf x}).\label{slotb}
\end{equation}
In this appendix we prove that for quadratic functions, and for
${\bf HA}$ with $p$ distinct eigenvalues, Eq.(\ref{slotb}) is
equal to the conjugate gradient expression after p steps, not
including the first guess.

First we prove Eq.(\ref{M}).  For {\bf M} with p distinct
eigenvalues we have
\begin{equation}
\prod_{j=1}^{p} (\epsilon_j - {\bf M})=0
\end{equation}
where $\epsilon_j$ are the distinct eigenvalue of ${\bf M}$.
Therefore the identity can be written
\begin{equation}
{\bf I} ={\bf I}-\beta\prod_{j=1}^p (\epsilon_j - {\bf M})
\end{equation}
for any $\beta$. Consequently
\begin{equation}
{\bf M}^{-1} = ({\bf I}-\beta\prod_{j=1}^p (\epsilon_j - {\bf
M})){\bf M}^{-1}
\end{equation}
for any $\beta$.  In general this is an implicit equation for
${\bf M}^{-1}$ but with $\beta = 1/\prod_{k=1}^p \epsilon_k$ each
term in parenthesis contains at least one power of ${\bf M}$,
cancelling out ${\bf M}^{-1}$ on the right hand side.  The final
result expresses ${\bf M}^{-1}$ explicitly as a polynomial in
${\bf M}$ alone with highest power $p-1$, as alleged in
Eq.(\ref{M}). The results for p from one to four are
\begin{eqnarray}
{\bf M}_1^{-1} &=& \frac{1}{\epsilon_1}{\bf I}\\ {\bf M}_2^{-1}
&=& \left(\frac{1}{\epsilon_1}+ \frac{1}{\epsilon_2}\right){\bf I}
-\frac{1}{\epsilon_1\epsilon_2}{\bf M}_2\\ {\bf M}_3^{-1}
&=&\left(\frac{1}{\epsilon_1}+\frac{1}{\epsilon_2}+\frac{1}{\epsilon_3}\right){\bf
I}
-
\left(\frac{1}{\epsilon_1\epsilon_2}+\frac{1}{\epsilon_1\epsilon_3}+\frac{1}{\epsilon_2\epsilon_3}
\right){\bf M}_3 + \frac{1}{\epsilon_1\epsilon_2\epsilon_3}{\bf
M}_3^2\\ {\bf M}_4^{-1} &=&
\left(\frac{1}{\epsilon_1}+\frac{1}{\epsilon_2}+\frac{1}{\epsilon_3}+\frac{1}{\epsilon_4}\right){\bf
I}
-\left(\frac{1}{\epsilon_1\epsilon_2}+\frac{1}{\epsilon_1\epsilon_3}+\frac{1}{\epsilon_1\epsilon_4}
+\frac{1}{\epsilon_2\epsilon_3}+\frac{1}{\epsilon_2\epsilon_4}+\frac{1}{\epsilon_3\epsilon_4}\right)
{\bf
M}_4\nonumber\\&+&\left(\frac{1}{\epsilon_1\epsilon_2\epsilon_3}+
\frac{1}{\epsilon_1\epsilon_2\epsilon_4}+
\frac{1}{\epsilon_1\epsilon_3\epsilon_4}+
\frac{1}{\epsilon_2\epsilon_3\epsilon_4}\right){\bf M}_4^2
-\frac{1}{\epsilon_1\epsilon_2\epsilon_3\epsilon_4}{\bf M}_4^3
\end{eqnarray}

Next we show that conjugate gradient minimization is equal to
Eq.(\ref{slotb}) after p steps. The proof is inductive.  We will
prove that for conjugate gradient minimization
\begin{equation}
{\bf x}_n = {\bf x}_1 + P_{n-2}({\bf HA}){\bf H}\nabla E({\bf
x}_1)\label{induct}
\end{equation}
where $P_{n-2}({\bf HA})$ is a polynomial of order $n-2$.  First,
it is trivially true for $n=2$ since ${\bf x}_2 = {\bf x}_1 -{\bf
H}\nabla E({\bf x}_1)$.  We will now prove that if it is true for
the pth step then it is also true for the p+1th step where
\begin{equation}
{\bf x}_{p+1} = {\bf x}_{p} -\lambda_p {\bf H}\cdot{\bf \nabla}
E({\bf x}_n) +\lambda_p\sum_{j=2}^{p}a_j({\bf x}_j-{\bf x}_{j-1}),
\end{equation}
in which the $a_j$'s are the coefficients of your favorite
conjugate gradient routine. Using Eq.(\ref{Arule}) we can put
things in terms of ${\bf x}_1$.
\begin{equation}
{\bf x}_{p+1}= {\bf x}_1 + ({\bf x}_p-{\bf x}_{1}) -\lambda_p{\bf
H}\cdot({\bf \nabla} ( E({\bf x}_1) + {\bf A}({\bf x}_p-{\bf
x}_{1})) +\lambda_p\sum_{j=2}^p a_j(({\bf x}_j-{\bf x}_{1}) -({\bf
x}_{j-1}-{\bf x}_{1}))
\end{equation}
Therefore using Eq.(\ref{induct}) we have
\begin{equation}
{\bf x}_{p+1} = {\bf x}_1 +\left(-\lambda_p {\bf HA}P_{p-2}({\bf
HA}) + P_{p-2}({\bf HA}) +\lambda_p\sum_{j=2}^p a_p( P_{j-2}({\bf
HA})-P_{j-3}(\bf{HA})) -\lambda_p\right)\cdot{\bf H}\cdot({\bf
\nabla} E({\bf x}_1)\label{taba},
\end{equation}
where $P_{-1}=0$.  The expression in parenthesis, because of its
first term, is a polynomial of order $p-1$.  Indeed it can be any
polynomial of order $p-1$ since $\lambda_p$ and the p-1 $a_j$'s in
principle can be varied arbitrarily.  In particular it can be the
polynomial in Eq.(\ref{slotb}).  The conjugate gradient formula
picks out the polynomial of order $p-1$ that minimizes $E({\bf
x}_{p+1})$. Since $E$ is minimized for ${\bf x}_{p+1}={\bf x}_{\rm
min}$, Eq.(\ref{taba}) must equal Eq.(\ref{slotb}) and the theorem
is proved.

We acknowledge support from Research Corporation grant CC5326.
R.A. Hyman thanks the hospitable folks of the James Frank
Institute of the University of Chicago where portions of this
paper were written.

\begin{references}
\bibitem{scooter} The scooter represents a routine that minimizes the function along a particular
ray. A real line minimization routine moves quite unlike the
motion of any real scooter, more like several pogo sticks bouncing
along the ray until the minimum is found. If there is more than
one minima, such routines can not guarantee that the minima found
is the closest one to your starting point.  References for line
minimization routines can be found the next citation.
\end{references}
\end{document}